\documentclass[11pt]{article}
\usepackage{booktabs} 
\usepackage{xcolor}
\usepackage{graphicx}
\usepackage{array}
\usepackage{setspace}
\usepackage{subcaption} 
\usepackage{listings}
\usepackage{caption}
\usepackage{float}
\usepackage{authblk}
\usepackage{tabularx}
\usepackage{comment}
\usepackage{enumitem}
\usepackage[most]{tcolorbox}
\usepackage{amsmath,amssymb}
\usepackage{xurl}
\newcommand{\leanid}[1]{\path{#1}}

\definecolor{torchblue}{RGB}{16,34,170}
\newtcolorbox{informaltheorem}[1][]{
  enhanced,
  breakable,
  colback=gray!8,
  colframe=torchblue,
  coltitle=white,
  fonttitle=\bfseries,
  title=#1,
  boxrule=1pt,
  arc=2pt,
  left=8pt,
  right=8pt,
  top=8pt,
  bottom=8pt
}

\usepackage{hyperref}
\definecolor{mycolor}{RGB}{0,88,204}
\hypersetup{
  colorlinks=true,
  linkcolor=mycolor,
  urlcolor=mycolor,
  citecolor=mycolor
}
\usepackage{amsmath}
\usepackage{geometry}

\usepackage{fancyhdr}

\fancypagestyle{plain}{%
  \fancyhf{}
  
}

\usepackage{mathptmx}

\usepackage{titlesec}

\titleformat{\section}
  {\normalfont\large\raggedright}{\thesection.}{1em}{}

\titleformat{\subsection}
  {\normalfont\raggedright}{\thesubsection.}{1em}{}

\titlespacing{\paragraph}{10pt}{0pt}{6pt}[0pt]

\usepackage{listings}
\usepackage[T1]{fontenc}
\usepackage[utf8]{inputenc}
\usepackage{amssymb}

\definecolor{keywordcolor}{rgb}{0.7, 0.1, 0.1}   
\definecolor{tacticcolor}{rgb}{0.0, 0.1, 0.6}    
\definecolor{commentcolor}{rgb}{0.4, 0.4, 0.4}   
\definecolor{symbolcolor}{rgb}{0.0, 0.1, 0.6}    
\definecolor{sortcolor}{rgb}{0.1, 0.5, 0.1}      
\definecolor{attributecolor}{rgb}{0.7, 0.1, 0.1} 

\lstnewenvironment{code}[1][]%
{
   \noindent\newline
   \minipage{\linewidth} 
   \vspace{0.5\baselineskip}
   \lstset{
 	frame = lrtb, 
 	rulecolor=\color{mycolor},
 	escapeinside={/*!}{!*/},
	language=lean, 
	aboveskip = 5mm,
	belowskip = 5mm,
	xleftmargin=2mm,
	xrightmargin=2mm,
	}
	}
{\endminipage\newline}
\lstnewenvironment{codeLong}[1][]%
{
   \lstset{
 	frame = lrtb, 
 	rulecolor=\color{mycolor},
 	escapeinside={/*!}{!*/},
	language=lean, 
	aboveskip = 5mm,
	belowskip = 5mm,
	xleftmargin=2mm,
	xrightmargin=2mm,
	}
	}
{}

\title{LeanBET: Formally-verified surface area calculations in Lean}
\author[1]{\normalsize Ejike D. Ugwuanyi}
\author[1]{\normalsize Colin T. Jones}
\author[1]{\normalsize John Velkey}
\author[1,2,*]{\normalsize  Tyler R. Josephson}

\affil[1]{Department of Chemical, Biochemical, and Environmental Engineering, \authorcr University of Maryland, Baltimore County, Baltimore, MD, USA}
\affil[2]{Department of Computer Science and Electrical Engineering, \authorcr University of Maryland, Baltimore County, Baltimore, MD, USA}
\affil[*]{Corresponding author, email: tjo@umbc.edu}

\begin{document}

\maketitle 
\begin{abstract}
The Brunauer--Emmett--Teller (BET) method is a standard tool for estimating surface areas from adsorption isotherms, yet practical implementations involve multiple algorithmic steps whose correctness is rarely made explicit. In this work, we present a fully executable and formally verified BET analysis pipeline implemented in the Lean~4 theorem prover.

Our formalization covers the complete BET Surface Identification (BETSI)-style workflow, including window enumeration, monotonicity checks, knee selection, and linear regression. We carry out computations in floating-point arithmetic and develop the corresponding correctness proofs over the real numbers, using a shared polymorphic implementation that supports both. On the proof side, we show that the regression coefficients returned by the algorithm agree with their specification-level definitions and minimize the least-squares error under the stated assumptions. We also formalize the algebraic derivation of the BET linearized expression and connect that result directly to the executable analysis pipeline. We further prove that the window enumeration is sound and complete, and that the admissibility checks and knee-based selection satisfy their formal specifications.

We evaluate the implementation against the BETSI reference method on benchmark adsorption isotherms. Compared to BETSI, LeanBET agrees to machine precision for 18 of the 19 isotherms, with only a 0.03\% deviation for the UiO-66 dataset. This demonstrates that a scientific computing workflow can be built in Lean, yielding both formal verification guarantees and numerical agreement with an established Python reference implementation.

\end{abstract}

\section{Introduction}
Scientific computing often relies on mathematical models whose derivations are written on paper, while their implementations are carried out separately in software. This separation can make it difficult to verify that the final computation faithfully reflects the underlying theory. Formal verification provides a way to address this problem. It is a method of checking, with machine assistance, that a computation follows from clearly stated mathematical definitions and logical steps. By expressing both the mathematics and the computation in a form that can be checked mechanically, formal verification helps bridge the gap between theory and implementation. In this work, we use derivation as specification. The algebraic development of the BET adsorption equation is formalized and directly linked to an executable analysis pipeline. We demonstrate this approach to scientific computing by formalizing the theory behind Brunauer–Emmett–Teller Theory and its implementation in surface area calculations.

The Brunauer–Emmett–Teller (BET) theory describes multilayer physical adsorption of gas molecules on solid surfaces and remains one of the most widely used frameworks for quantifying specific surface area from adsorption isotherms\cite{brunauer_adsorption_1938}.
In practice, BET analysis is usually carried out by selecting a relative pressure range, fitting the linearized BET equation over that interval, and extracting the monolayer capacity and surface area from the resulting regression parameters. Although the method is standard, the final surface area can depend strongly on the chosen fitting range. As a result, BET calculations that begin from the same isotherm may still lead to different reported values, reducing reproducibility across users, laboratories, and software implementations\cite{van_erp_standardization_2011}\cite{osterrieth_how_2022}.

Despite its longevity and prevalence, BET surface area analysis has long been known to rest on a set of restrictive assumptions about adsorption behavior\cite{ambroz_evaluation_2018}. The BET model extends the Langmuir monolayer adsorption framework to multilayer adsorption, assuming uniform surface sites, non-interacting adsorbate layers beyond the first, and a constant enthalpy of adsorption for layers above the monolayer\cite{brunauer_adsorption_1938}. These assumptions only hold approximately for many real materials, and the BET plot often exhibits linearity over a narrow and sometimes ambiguous relative-pressure range; hence, selecting an appropriate fitting region has become a central practical challenge in BET analysis\cite{do_computer_2010}. To mitigate this ambiguity, a set of consistency criteria frequently attributed to Rouquerol and co-workers and adopted in practice has been used to help identify valid regions of linearity in BET plots and to exclude unphysical solutions such as negative monolayer capacities. These criteria include requiring a positive BET constant \(C\), a monotonically increasing \(n(1-P/P_0)\) term over the selected range, and that the calculated monolayer coverage corresponds to a relative pressure within that range \cite{rouquerol_recommendations_1994}. However, even when such criteria are applied, additional subjectivity remains, because multiple relative pressure intervals may satisfy the same consistency rules. This leaves the lower and upper bounds of the fitting range unspecified, leading to variability in reported BET areas.

This subjectivity has profound consequences. In a recent round-robin study involving 61 laboratories and a shared set of high-quality adsorption isotherms, Osterrieth \emph{et al.} demonstrated that BET surface areas calculated from the same data varied widely across laboratories, even when most participants reported using the consistency criteria in their analysis \cite{osterrieth_how_2022}. For some materials, the highest and lowest reported BET areas differed by more than a factor of five, highlighting that reproducibility issues arise not from experimental measurement alone, but also the fitting algorithms and human choices used to interpret that data. To address this reproducibility crisis, Osterrieth \emph{et al.} introduced BET Surface Identification (BETSI), a Python-based computational framework that exhaustively enumerates all consecutive fitting windows on an isotherm, applies ordinary least-squares regression to each, and filters the resulting candidates using the Rouquerol criteria to produce an unambiguous BET area assignment \cite{osterrieth_how_2022}. While BETSI substantially reduces the subjectivity associated with manual fitting, it remains a numerical algorithm implemented in general-purpose software, leaving open the possibility of hidden implementation assumptions or subtle errors that could compromise its scientific guarantees.

\begin{figure}[t]
  \centering
  \includegraphics[width=\textwidth]{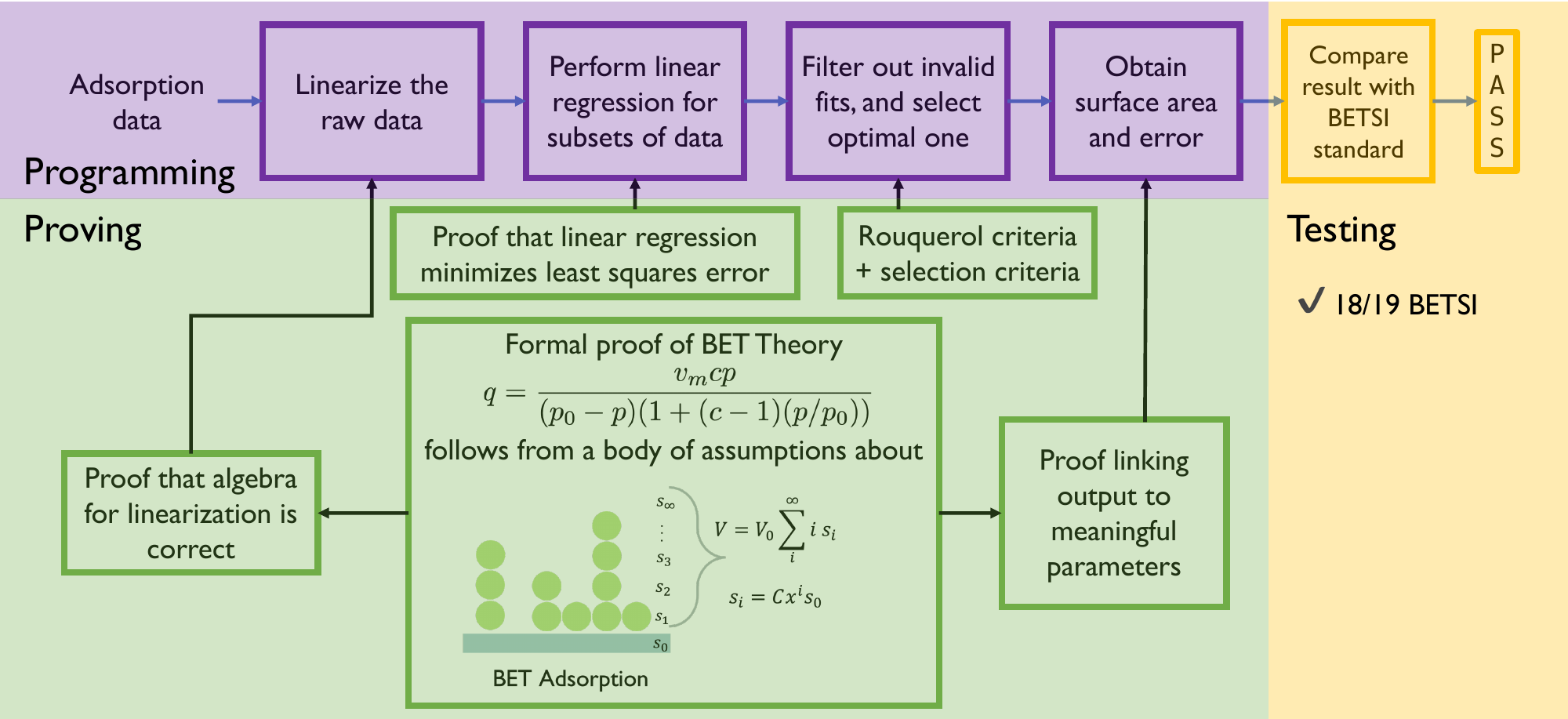}
  \caption{
  \textbf{Overview of the formally verified BET analysis pipeline.}
  The executable algorithm (top row) transforms adsorption data through BET linearization, windowed linear regression, and BETSI-style filtering to compute surface area and error metrics.
  Each computational step is paired with a corresponding formal proof (bottom blocks) establishing algebraic correctness of the BET linearization, least-squares optimality of linear regression, and soundness of the selection criteria.
  Numerical results are finally compared against the BETSI reference implementation.
  }
  \label{fig:bet_formal_pipeline}
\end{figure}

In this work, we present a different approach to BETSI-style analysis. Figure~\ref{fig:bet_formal_pipeline} provides a schematic overview of the formally verified BET analysis pipeline, illustrating how each executable transformation is justified by a corresponding mathematical proof. Rather than focusing solely on numerical performance or agreement with reference values, we develop a \emph{verified} BETSI implementation in which correctness properties are formally specified and mechanically verified. The algorithm is implemented in Lean~4 which is both a programming language and theorem proving system that allows executable code and mathematical proofs to be written in the same environment \cite{de_moura_lean_2021}. When the algorithm returns a BET fit, Lean provides a machine-checked proof that the result satisfies the BETSI criteria as mathematically defined. The central contribution of this paper is therefore not a new BET algorithm, but a formally verified guarantee of correctness. The implementation is designed to be polymorphic, supporting efficient execution using floating-point arithmetic while enabling proofs over the real numbers. By separating computation from specification and proving soundness of the executable checks, we demonstrate how formal methods can be applied to a classical problem in adsorption science without sacrificing practical usability.

The remainder of this paper is organized as follows. Section~\ref{sec:background} reviews the theoretical background of BET analysis and the BETSI procedure. Section~\ref{sec:Lean} introduces Lean 4 and outlines the design principles used to integrate executable code with formal verification. Section~\ref{sec:BETSI} describes the polymorphic implementation of BETSI. Section~\ref{sec:spec-and-soundness} presents the formal specifications and verified soundness of the BETSI algorithm. Section~\ref{sec:discussion} reports numerical results and comparisons with existing implementations. Finally, Section~\ref{sec:conclusion} discusses the conclusions and outlook.

\section{Background}
\label{sec:background}
\subsection{BET Theory and Linearization}

The Brunauer--Emmett--Teller (BET) method provides a classical framework for 
estimating the specific surface area of a solid from gas adsorption 
isotherms~\cite{brunauer_adsorption_1938}. The method extends the Langmuir monolayer 
adsorption model~\cite{langmuir_adsorption_1918} to multilayer adsorption by assuming 
that adsorption occurs in successive layers and that, beyond the first layer, the 
heat of adsorption is equal to the heat of liquefaction of the adsorbate. Bobbin et al.\cite{bobbin_formalizing_2024} were the first to rigorously formalize its derivation, encoding it as a machine-verified proof in the Lean~3 theorem prover, an achievement that clarified precisely which definitions are necessary and sufficient to imply the adsorption isotherm equation. The model is built on six definitions:
\begin{itemize}
  \item The first-layer adsorption rate $y = P \cdot C_1$,
  \item The multilayer adsorption rate $x = P \cdot C_L$,
  \item The BET constant $C = y/x$,
  \item The geometric layer occupancy $s_i = Cx^i s_0$,
  \item The total adsorbed area $A = \sum_{i=0}^{\infty} s_i$, and
  \item The total adsorbed volume $V = V_0 \sum_{i=0}^{\infty} i \cdot s_i$.
\end{itemize}
These definitions require the convergence condition $0 < x < 1$, a constraint 
absent from the original 1938 derivation but essential to the validity of both 
infinite sums. Under this condition, both series admit closed forms, and their 
ratio $V/A$ yields an expression in $x$ and $C$. Recognising that $x = P \cdot C_L$ 
is simply the relative pressure $P/P_0$ when the saturation pressure is identified 
as $P_0 = 1/C_L$, the ratio reduces to the standard nonlinear BET isotherm,
\begin{equation}
  n(P) = \frac{C \cdot n_m \cdot (P/P_0)}
              {(1 - P/P_0)\bigl(1 + (C-1)(P/P_0)\bigr)},
  \label{eq:BET_isotherm}
\end{equation}
where $p = P/P_0$ is the relative pressure, $n_m$ is the monolayer capacity, and 
$C$ is the adsorption energy constant. The formal derivation of 
Equation~\eqref{eq:BET_isotherm} from these definitions is presented as a 
machine-checked theorem in Section~5.3.

Rearranging Equation~\eqref{eq:BET_isotherm} into an affine form yields the 
linearized BET equation,
\begin{equation}
  \frac{p}{n(1-p)} = \frac{C-1}{n_m C}\, p + \frac{1}{n_m C}.
  \label{eq:BET_linear}
\end{equation}
When the BET equation is valid, a plot of $p/[n(1-p)]$ versus $p$ reveals a linear 
region from which $n_m$ and $C$ can be recovered by linear regression. Despite the 
apparent simplicity of this procedure, practical BET analysis is critically dependent 
on the identification of an appropriate pressure range over which 
Equation~\eqref{eq:BET_linear} holds.

\subsection{Rouquerol Criteria}

To reduce ambiguity in the choice of the BET fitting range, the
analysis applies a sequence of self-consistency checks to each
candidate pressure window~\cite{rouquerol_recommendations_1994}.
The four criteria are as follows:

\begin{enumerate}[label=(\roman*)]
    \item The quantity $n(1 - P/P_{0})$ must be monotonically non-decreasing over the fitting window.

    \item The fitted BET parameters must be physically admissible:
          $C > 0$ and $n_{m} > 0$.

    \item The relative pressure corresponding to monolayer coverage,
          $P/P_{0}\big|_{n_{m}}$, must lie within the selected
          fitting window.

    \item The monolayer pressure predicted analytically by the fitted
          model, $P/P_{0} = 1/(\sqrt{C}+1)$, must agree with the
          value read from the isotherm to within a prescribed
          tolerance of 20\%.
\end{enumerate}

A candidate window is considered valid only if all four
criteria are satisfied simultaneously.
\subsection{BETSI Algorithm}

The BET Surface Identification (BETSI) approach was developed to address the reproducibility challenges inherent in BET analysis by providing a fully algorithmic procedure for window selection and evaluation \cite{osterrieth_how_2022}. Rather than relying on manual inspection or subjective judgment, BETSI systematically enumerates all admissible pressure windows within an isotherm and applies a sequence of physical and statistical filters to identify valid BET regions. At a high level, the BETSI procedure consists of the following steps. First, all contiguous pressure windows containing at least two data points are generated. For each window, the BET equation is linearized and a least-squares regression is performed. Windows that fail to satisfy the Rouquerol criteria or that yield physically inadmissible parameters are discarded. For the remaining candidates, the BET equation is inverted to estimate the pressure corresponding to monolayer coverage, and a relative error criterion is applied. Finally, among the passing windows, a knee selection strategy is used to identify a single representative BET fit.

This approach adds an additional criterion to the four from Rouquerol, to admit a unique solution:
\begin{enumerate}[label=(\roman*), resume]
    \item If several windows satisfy the previous conditions, the final result is chosen from those reaching the largest admissible end index, with the percentage error used to break any remaining ties.
\end{enumerate}

The strength of the BETSI approach lies in its systematic nature and its ability to remove user-dependent choices from BET analysis. However, as with any numerical algorithm, its correctness ultimately depends on the faithful implementation of each step and on the interpretation of the underlying criteria. The work presented here builds on the BETSI methodology by providing a formally verified implementation in which these correctness properties are explicitly specified and mechanically checked.


\section{Lean 4 Framework and Design Approach}
\label{sec:Lean}

\subsection{Lean 4 as the Formalization Tool}
The formal verification work presented in this paper is carried out using the Lean 4 theorem prover \cite{de_moura_lean_2021}.
Lean~4 is not only a theorem prover but also an efficient functional programming language, which makes it possible to develop executable code and machine-checked proofs within the same environment \cite{Christiansen_functional_programming_lean}. This dual role has motivated a growing body of work on verified computation in Lean. For example, SciLean explores scientific computing workflows in Lean \cite{skrivan_lecopivoscilean_2025}, while TorchLean extends this idea to neural-network models and verification \cite{george_torchlean_2026}. Earlier, Selsam et al.\cite{selsam_developing_2017} demonstrated that Lean could be used to implement and verify core properties of a machine learning system. Our work follows this general direction, but focuses on adsorption analysis, using Lean to connect the derivation, implementation, and verification of BET-style surface area computation.

All results reported in this work were developed using Lean~4, together with the accompanying mathematical library \texttt{Mathlib v4.27.0-rc1}. The Lean build system \texttt{lake} was used to manage dependencies and compilation. The specific Lean version used for this work is Lean~4.27.0, although the design does not rely on features unique to this release.

A key advantage of Lean for scientific computing applications is that executable code and mathematical specifications can be expressed side by side. Numerical algorithms can be written in a style similar to conventional functional programming, while correctness properties are stated as logical propositions and proved within the same environment. This avoids the need to translate algorithms into a separate verification language and reduces the risk of mismatches between code and specification \cite{lean_capability_mateo}\cite{programming_in_lean_avigad_hudon}.

\subsection{Polymorphic Numeric Design}
\begin{figure}
    \centering
    \includegraphics[width=0.5\linewidth]{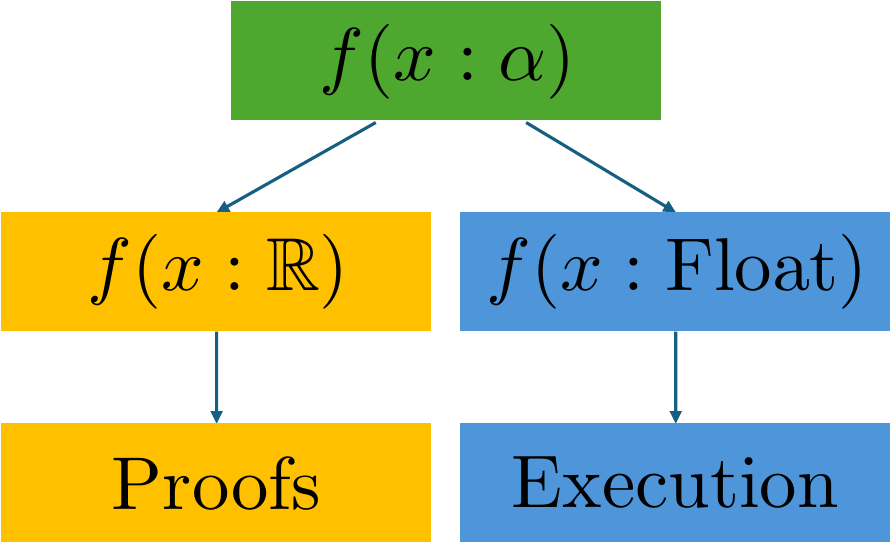}
    \caption{Polymorphic functions link proofs (over idealized Real numbers) with execution (over floating point numbers), using a polymorphic function $f$ that uses generic type $\alpha$. To set up proofs, $\alpha$ is specialized to Real, and to set up executions, $\alpha$ is specialized to Float. Figure taken from \cite{ugwuanyi_benchmarking_2025} with permission.}
    \label{fig:polymorphism}
\end{figure}

A central challenge in this formalization is that the two roles we need from a numeric representation are not available in a single type. Floating-point numbers are executable and therefore suitable for running the BETSI pipeline on real data, but they are not the basis of the mathematical proofs developed here. Real numbers, by contrast, provide the natural setting for specification and proof, but in Lean they are \emph{noncomputable} and therefore not suitable for direct execution of the full analysis. Our solution is to express the BETSI algorithm using polymorphic types, so that the same algorithmic structure can be instantiated over floating-point values for computation and over real numbers for proof (Figure~\ref{fig:polymorphism}). In this way, polymorphism provides the mechanism that links executable computation with proof-oriented specification. Thus, we prove correctness for functions over real numbers, and perform execution for functions over floating-point values. This approach differs from methods that reason directly about floating-point operations, including formally verified floating-point libraries such as Flocq \cite{boldo_flocq_2011} and tools for automated round-off error certification such as Gappa \cite{de_dinechin_certifying_2008}. Within the Lean ecosystem, related efforts include experimental developments such as Flean \cite{mckinsey_flean_2025}, which aim to formalize floating-point semantics directly. It also contrasts with approaches based on interval arithmetic, which provide certified bounds on numerical error (e.g., CoqInterval \cite{melquiond_floating-point_2012} and LeanCert \cite{alejandro_leancert_2026}).

In our implementation, we handle the alignment between Real- and Float-valued functions using the typeclass \texttt{BETLike}, which organizes the instances for the arithmetic operations used in all the functions in this work. Unlike the \texttt{RealLike} typeclass from LeanLJ \cite{ugwuanyi_benchmarking_2025}, which uses Boolean-valued comparisons suited for computation, \texttt{BETLike} uses proposition-valued ordering that integrates naturally with Lean's mathematical proof infrastructure.

\subsection{Executable Checks and Derivation as Specification}

Specifications in this work come from two distinct sources. The first is the set of mathematical properties that each function is expected to 
satisfy. The monotonicity check corresponds to a sequence being nondecreasing, the regression step to solving a least-squares problem, and 
the window enumeration to producing every valid contiguous sublist of the isotherm. For each of these, we prove the property holds whenever the 
corresponding executable check succeeds, one function at a time.

The second source is the BET derivation itself. Working through the algebra of the BET model is what produces the linearized form $p/[n(1-p)]$ at the center of the pipeline. The derivation does not just motivate this transformation; it implies it, and in doing so fixes what the code must 
compute. This is what we mean by derivation as specification.

One further point is worth stating clearly. The correctness established here is a property of the code, not of any particular run. Once the proofs 
compile, every execution inherits that guarantee automatically. A returned result carries evidence that it passed all the checks, but that evidence is trustworthy because the code was proved correct beforehand, not because the individual run was validated after the fact.
\section{Implementation of LeanBET}
\label{sec:BETSI}

This section describes the structure of our implementation of the BETSI algorithm. The presentation focuses on how the algorithm operates on adsorption data. All components are written in a polymorphic style, allowing the same implementation to be used for floating-point execution and real-valued reasoning. The full source is available at \url{https://github.com/ATOMSLab/LeanBET}. In Section~\ref{sec:spec-and-soundness}, we highlight the proofs verifying the functions described here.

\subsection{Data Structures}

The core object used throughout the analysis is the adsorption isotherm, represented as a finite ordered list of data points. Each point contains a relative pressure value and the corresponding uptake, each with type $\alpha$ (for proofs, we specialize $\alpha$ to Real, for execution, we specialize $\alpha$ to Float). This provides the input for all later fitting and filtering steps.

\begin{code}
structure Point (α : Type) where
  p : α -- reduced pressure
  n : α -- amount adsorbed
  deriving Repr

abbrev Isotherm (α : Type) := List (Point α)
\end{code}

Additional structures are used to store intermediate and final results. A fit record \texttt{BETFit} stores the slope and intercept of the linearized BET regression, the coefficient of determination, the derived monolayer capacity and BET constant, and the index range of the fitting window. A final result record \texttt{BETSIResult} also stores the selected window together with auxiliary quantities used in the last stage of filtering, such as the pressure values and percentage error. Keeping these quantities explicit makes the procedure easier to inspect, easier to verify, and easier to compare across candidate windows. Variables in these either have type $\alpha$ or type \texttt{Nat} (for the natural numbers) -- since natural numbers are computable, we don't need to invoke polymorphism for them.

\begin{code}
structure BETFit (α : Type) where
  slope : α
  intercept : α
  rSquared : α
  nm : α
  c : α
  range : Nat × Nat
  nPoints : Nat
  deriving Repr

structure BETSIResult (α : Type) where
  fit     : BETFit α
  window  : List (Point α)
  p_nm    : α
  p_read  : α
  pcError : α
  deriving Repr
\end{code}

\subsection{BET Linearization}
The next step applies the standard BET linearization to each data point in the isotherm. Here \(p\) denotes the relative pressure, defined as \( p = P/P_0 \), and \(n\) denotes the corresponding uptake. In the present implementation, the isotherm is represented directly in terms of relative pressure and uptake, so the transformation is applied to reduced pressure values.

Following the linearized BET form introduced earlier in Eq.~\eqref{eq:BET_linear}, each point is mapped to the pair $\left(p,\; \dfrac{p}{n(1 - p)}\right)$. This is the quantity used in the regression step. The transformation is applied only when the denominator \(n(1-p)\) is defined and nonzero, thus avoiding division by zero. In this way, the executable transformation mirrors the algebraic linearization developed in Section~\ref{sec:background}.

\subsection{Window Enumeration}
BET analysis is carried out over contiguous subsets of the isotherm, which we refer to as windows. The implementation enumerates every contiguous window containing at least two points and records its start index, end index, and associated sublist of data. Each such window is then passed to the regression and admissibility steps described below.

This gives a finite collection of candidate fitting ranges drawn from the experimental isotherm. By storing the window boundaries explicitly, the later stages of the algorithm can track which interval produced each candidate fit and result.

\subsection{Linear Regression}
For each window, the transformed data are fitted by least-squares linear regression, according to the linearized BET relation in Eq.~\eqref{eq:BET_linear}. The regression operates on a list of pairs $(x_i, y_i)$, where $x_i = p_i$ is the relative pressure and $y_i = p_i / [n_i(1 - p_i)]$ is the BET-transformed uptake.

\begin{code}
/-- Least-squares linear regression returns (slope, intercept, r²). -/
def linearRegression {α} [BETLike α] (data : List (α × α)) : Option (α × α × α) := do
  if data.length < 2 then none else
    let n : α := BETLike.ofNat data.length
    let xs : List α := data.map Prod.fst
    let ys : List α := data.map Prod.snd
    let xBar : α := listSum xs / n
    let yBar : α := listSum ys / n
    let cov : α := listSum <| data.map (fun (x, y) => (x - xBar) * (y - yBar))
    let vr : α := listSum <| data.map (fun (x, _) => (x - xBar) ^ (2 : Nat))
    
    if BET.isZero vr then none else
      let slope : α := cov / vr
      let intercept : α := yBar - slope * xBar
      let yHat : List α := xs.map (fun x => slope * x + intercept)
      let ssTot : α := listSum <| ys.map (fun y => (y - yBar) ^ (2 : Nat))
      let ssRes : α := listSum <| (List.zip ys yHat).map (fun (y, yh) => (y - yh) ^ (2 : Nat))
      let r2 : α := if BET.isZero ssTot then BETLike.one else BETLike.one - (ssRes / ssTot)
      some (slope, intercept, r2)
\end{code} 

Given $n$ such pairs, the slope and intercept of the fitted line are
\begin{equation}
  m = \frac{\sum_{i=1}^{n}(x_i - \bar{x})(y_i - \bar{y})}
           {\sum_{i=1}^{n}(x_i - \bar{x})^2}, 
  \qquad
  b = \bar{y} - m\bar{x},
  \label{eq:regression}
\end{equation}
where $\bar{x}$ and $\bar{y}$ are the sample means. The goodness of fit is assessed 
via the coefficient of determination,
\begin{equation}
  R^2 = 1 - \frac{\sum_{i=1}^{n}(y_i - \hat{y}_i)^2}
                 {\sum_{i=1}^{n}(y_i - \bar{y})^2},
  \label{eq:r2}
\end{equation}
where $\hat{y}_i = mx_i + b$ is the fitted value at $x_i$. From the returned slope 
and intercept, the monolayer capacity and BET constant are recovered via
\begin{equation}
  n_m = \frac{1}{b + m}, \qquad C = 1 + \frac{m}{b}.
  \label{eq:BET_params}
\end{equation}
The function returns \texttt{none} if fewer than two points are provided or if the 
variance of the $x_i$ values is zero, since the regression is undefined in those 
cases.

\subsection{Admissibility Checks}
Each candidate fit is then tested against the Rouquerol admissibility criteria described in Section~2. These checks include monotonicity conditions on derived adsorption quantities, minimum data requirements, coefficient of determination, and basic physical constraints such as positivity of the fitted parameters. A candidate is retained only if it satisfies all of the required conditions. The outcome is a filtered collection of admissible candidate fits, which are then passed to the final selection stage.

\subsection{Final Selection}
The Rouquerol criteria alone do not guarantee a unique result, and when multiple candidate windows pass all four conditions, a fifth criterion is needed. BETSI supplies this through a knee-selection rule, and in LeanBET this rule is implemented as a two-stage function over the polymorphic type \texttt{BETLike}~$\alpha$.

The first stage filters the list of passing candidates by end index, retaining only those whose fitting window reaches the largest pressure index among all retained candidates. This is implemented as a list traversal that compares end indices using the \texttt{BETLike} ordering, and returns a sublist. The second stage selects from that sublist the candidate with the smallest percentage error, implemented as a fold over the filtered list.

The correctness of the selection, that the returned result genuinely belongs to the maximal-end-index candidates is not assumed but proved, as stated in Theorems~A.6 and A.7. Together with the Rouquerol checks this completes the LeanBET implementation of the full BETSI procedure.

\section{Formal Specification and Verified Soundness of the BETSI Algorithm}
\label{sec:spec-and-soundness}

This section presents the formal interpretation of the BETSI algorithm together with the correctness guarantees established for its implementation. The goal is to make precise what it means for a BETSI result to be valid and to show that every result produced by the algorithm satisfies this interpretation. All specifications and proofs are carried out in Lean~4, using a shared polymorphic formulation that supports both floating-point execution and real-number reasoning. Throughout this section, key results are stated as informal theorems in natural language, labeled Theorem~A.1 through Theorem~A.7; the complete machine-checked Lean proofs and supporting lemmas are available in the accompanying repository.

\subsection{Specification of the BETSI-style Procedure}

The BETSI-style procedure considered in this work consists of a sequence of conceptual steps: enumerating admissible windows of an adsorption isotherm, fitting linearized BET models to those windows, applying physical and statistical admissibility criteria, and selecting a final result based on the knee principle. In this development, each of these steps is given an explicit mathematical specification, independent of any particular numerical implementation. The derivation also plays an important role here: it justifies the linearized BET form used in the regression step and helps connect the scientific model to the executable analysis.

Window enumeration is specified as the collection of all contiguous sublists of the isotherm containing at least two data points, together with their corresponding index ranges. This makes precise the search space over which BET fitting is carried out and ensures that no admissible window is omitted by design.

Monotonicity conditions, such as those appearing in the extended Rouquerol criteria, are specified as properties of derived sequences. For example, the sequence \(n(1-p)\) associated with a window is required to be nondecreasing, and the sequence obtained from BET linearization must satisfy the same property.

The regression step is specified in terms of least-squares fitting. Given a list of linearized data points, a valid regression result is defined as one whose slope and intercept minimize the sum of squared residuals, provided that the variance of the independent variable is nonzero. From these regression parameters, the BET constants \(n_m\) and \(C\) are defined using their standard theoretical formulas.

Finally, the knee-selection principle is specified as follows: among all admissible candidate results, those whose fitting window extends to the largest pressure index are retained, and from this subset the result with the smallest percentage error is selected. This captures the intended physical interpretation of the knee as the upper limit of reliable BET behavior.

Taken together, these specifications define what it means for a reported result to be valid, independently of how that result is computed.

\subsection{Executable Checks and Soundness}

The algorithm realizes these specifications through executable checks on numerical data. In practice, these checks are written so they can run efficiently on floating-point inputs. For example, monotonicity is checked by scanning finite lists, and the admissibility conditions are implemented as decision steps in the code.

The main correctness goal is \emph{soundness}: whenever one of these checks succeeds and the algorithm returns a result, the corresponding mathematical specification is satisfied. In other words, every reported result meets the formal validity conditions defined for the BETSI-style procedure.

The executable pipeline runs with floating-point arithmetic, while the main correctness statements are proved over the real numbers. As with any floating-point implementation, this leaves a small possibility that a comparison very close to a threshold could behave differently in execution than it would over exact real values. In practice, however, this is not a serious concern for the adsorption datasets considered here, since the relevant inequalities are not usually decided at machine-precision boundaries (for instance, 0.1 + 0.2 > 0.3 is true for IEEE-754 floats, but only deviates at the 17th decimal place). For this application, experimental noise is far greater than double-precision round-off values. 

This design places the emphasis on the correctness of accepted results. The algorithm enforces the BETSI-style admissibility conditions and rules out outputs that are physically inadmissible or internally inconsistent. As a result, each surface area reported by the implementation comes with a machine-checked guarantee that it satisfies the formal specification.

\subsection{Soundness of the Linearized BET Model}

The formal guarantees for the linearization step rest on two connected results. 
The first shows that the BET isotherm (Eq.~\eqref{eq:BET_isotherm}) is a theorem derived from the layer
model, not an assumption. The second shows that the executable \texttt{linearizeBET} 
function, fed this proved isotherm, returns exactly the affine BET line.

\subsubsection*{Derivation of the BET Isotherm from the Layer Model}

The infinite sums for $A$ and $V$ defined in Section~2.1 both converge when 
$0 < x < 1$. Under this condition, evaluating their closed forms and taking the 
ratio $V/A$ gives the \href{https://github.com/ATOMSLab/LeanBET/blob/be6762163b6c8b35f7aa02d967219ea0ffb4b4e6/BET/BET_Infinite.lean#L86}{theorem} \leanid{brunauer_26_from_seq} intermediate result,
\begin{equation}
  \frac{V}{V_0 \cdot A} = \frac{Cx}{(1-x)(1 - x + Cx)}.
\end{equation}
Substituting $x = P/P_0$ then gives the standard BET isotherm, which is machine-checked in \href{https://github.com/ATOMSLab/LeanBET/blob/be6762163b6c8b35f7aa02d967219ea0ffb4b4e6/BET/BET_Infinite.lean#L277}{theorem} \leanid{brunauer_28_from_seq}. This means that the resulting equation follows from the layer definitions whenever the convergence condition holds, rather than being stated as a hypothesis.

\begin{informaltheorem}[Theorem A.1: The BET isotherm follows from the layer model]
Let $S$ be a BET system with saturation pressure $P_0 = 1/C_L$, and let $P > 0$ 
satisfy the convergence condition $0 < x < 1$, where $x = P \cdot C_L$. Then the 
ratio of total adsorbed volume to total adsorbed area satisfies
\[
  \frac{V}{A} = V_0 \cdot n(P),
\]
where $n(P)$ is the standard BET isotherm
\[
  n(P) = \frac{C \cdot n_m \cdot (P/P_0)}
              {(1 - P/P_0)\bigl(1 + (C-1)(P/P_0)\bigr)}.
\]
Thus, whenever the convergence condition holds and $P_0 = 1/C_L$, the standard 
BET isotherm is a direct consequence of the infinite layer definitions for $A$ and $V$.
\end{informaltheorem}

\newpage 
\subsubsection*{Algebraic Correctness of the Linearization}

Given the proved isotherm, the following theorem confirms that packaging $n(P)$ 
into the executable \href{https://github.com/ATOMSLab/LeanBET/blob/be6762163b6c8b35f7aa02d967219ea0ffb4b4e6/BET/Function.lean#L87}{def} \leanid{linearizeBET} function returns exactly the affine BET 
line, with slope and intercept encoding $n_m$ and $C$.

\begin{informaltheorem}[Theorem A.2: BET linearization is algebraically correct]
Given that the adsorption amount satisfies the nonlinear BET equation and $p=P/P_0$
\[
n(p)=\frac{C n_m p}{(1-p)\bigl(1+(C-1)p\bigr)},
\]
with \(C \neq 0\), \(n_m \neq 0\), \(p \neq 0\), and \(p \neq 1\). Then the BET-transformed quantity
\[
\frac{p}{n(p)(1-p)}
\]
is affine in \(p\), namely
\[
\frac{p}{n(p)(1-p)}
=
\frac{1}{C n_m}
+
\frac{C-1}{C n_m}\,p.
\]
Thus, whenever the nonlinear BET model holds, the standard BET plot is exactly linear, with intercept \(1/(C n_m)\) and slope \((C-1)/(C n_m)\).
\end{informaltheorem}

This \href{https://github.com/ATOMSLab/LeanBET/blob/main/BET/Linearization.lean}{theorem}
is the formal counterpart of the derivation presented earlier in Section 2. It shows that the quantity computed by the linearization step in the executable analysis is precisely the one predicted by BET theory.

\subsection{Soundness and Completeness of Window Enumeration}

The window-enumeration step defines the search space over which BET fitting is carried out. In the executable pipeline, a window is generated by choosing a start index, an end index, and returning the corresponding contiguous slice of the isotherm. For the later stages of the analysis to be meaningful, this enumeration step must do exactly what it is intended to do: it must generate only valid candidate windows, and it must generate all of them.

Formally, a valid window is a contiguous sublist of the original isotherm whose start and end indices lie within bounds and satisfy \(i<j\), so that the window contains at least two data points. The verification of this stage therefore has two parts. First, we show that every window returned by the executable procedure satisfies these validity conditions. Second, we show that every valid contiguous window determined by admissible indices is in fact returned by the procedure.

\newpage

\begin{informaltheorem}[Theorem A.3: Window enumeration returns exactly the valid contiguous fitting windows]
Let $I = [p_0, \ldots, p_{N-1}]$ be an input isotherm, and let $W = I[i:j+1] = [p_i, p_{i+1}, \ldots, p_j]$ denote the contiguous sublist of $I$ from index $i$ to index $j$ inclusive.
\textbf{Soundness.} Every triple \((i,j,W)\) returned by the window-enumeration procedure satisfies
\[
0 \le i < j < N
\qquad\text{and}\qquad
W = I[i:j+1].
\]

\textbf{Completeness.} Conversely, for every pair of indices \(i,j\) satisfying \[0 \le i < j < N,\]
the corresponding contiguous sublist \[W = I[i:j+1]\]
appears among the windows returned by the procedure.

Hence the executable enumeration procedure produces exactly the valid contiguous fitting windows, and each such window contains at least two points.
\end{informaltheorem}

This \href{https://github.com/ATOMSLab/LeanBET/blob/main/BET/WindowSoundness.lean}{theorem} shows that the fitting stage neither misses admissible candidate ranges nor introduces malformed ones. This supports the later regression, admissibility, and selection theorems, which can be stated over the intended BET search space itself.

Note that a Python implementation of BETSI could in principle ensure soundness by implementing Boolean if-then conditions to filter out invalid windows. However, Python does not have the machinery to assert completeness of the procedure; our implementation proves that we miss no windows.

\subsection{Soundness of Linear Regression and BET Parameter Extraction}

A successful regression means that the executable routine returns values $m$, $b$, and $r^2$. 
To state correctness precisely, we compare this result to an arbitrary alternative pair of coefficients $(m', b')$, representing any other candidate line.

\begin{informaltheorem}[Theorem A.4: Successful regression returns a least-squares minimizer]
Let \(D=\{(x_i,y_i)\}_{i=1}^n\) be a nonempty dataset of linearized adsorption points, and suppose the variance of the \(x_i\) values is positive. If the executable regression routine returns
\[
\mathrm{linearRegression}(D)=\mathrm{some}(m,b,r^2),
\]
then for every alternative pair of coefficients \((m',b')\),
\[
\mathrm{SS}_{\mathrm{res}}(D,m,b)
\;\le\;
\mathrm{SS}_{\mathrm{res}}(D,m',b').
\]
That is, the returned slope $m$ and intercept $b$ define a least-squares minimizer.
\end{informaltheorem}

Once these coefficients are obtained, the monolayer capacity $n_m$ and BET constant $C$ are computed directly from $m$ and $b$ using the standard formulas.

\begin{informaltheorem}[Theorem A.5: BET parameter extraction agrees with the certified regression line]
Suppose the executable regression routine returns
\[
\mathrm{linearRegression}(D)=\mathrm{some}(m,b,r^2).
\]
Then the quantities extracted from the returned line,
\[
n_m=\frac{1}{b+m}
\qquad\text{and}\qquad
C=1+\frac{m}{b},
\]
agree with their definitions from the linearized BET model.
\end{informaltheorem}

Together, these \href{https://github.com/ATOMSLab/LeanBET/blob/main/BET/RegressionSoundness.lean}{theorems} show that the regression stage correctly computes the slope and intercept of the fitted line, and that the physical parameters derived from this line are consistent with the underlying model. Parts of the formal development were interactively refined using proof-assistant tooling.

\subsection{Soundness of Admissibility Checks and Final Selection}

After regression, each candidate window is tested against the admissibility conditions derived from the Rouquerol criteria together with additional constraints used in the BETSI-style procedure. These include monotonicity of derived quantities, positivity of fitted parameters, and consistency conditions relating the monolayer pressure to the fitting range.

In the implementation, these conditions are evaluated through executable checks on finite sequences and fitted parameters. Each check returns a Boolean value indicating whether the condition is satisfied.

\begin{informaltheorem}[Theorem A.6: Soundness of admissibility checks]
Let $Z$ be a candidate window, and suppose all admissibility checks succeed for $Z$. Then the corresponding derived quantities satisfy the formal conditions: the relevant sequences are nondecreasing, the fitted parameters satisfy $C > 0$ and $n_m > 0$, and the monolayer pressure lies within the fitting range and meets the prescribed consistency condition.

Thus, any window accepted by the algorithm satisfies the full set of admissibility requirements.
\end{informaltheorem}

After filtering, the algorithm selects a single result using the knee-based rule. First, it determines the largest end index among all admissible windows. Then it restricts attention to those windows whose end index matches this maximum. Among these, the result with the smallest percentage error is selected.

\newpage

\begin{informaltheorem}[Theorem A.7: Soundness of knee-based selection]
Let $R$ be the set of admissible candidate results, and let $j_{\max}$ be the largest end index among them. If the selection procedure returns a result $r$, then:
\begin{itemize}
    \item $r \in R$,
    \item the fitting window of $r$ ends at $j_{\max}$, and
    \item among all candidates in $R$ with end index $j_{\max}$, $r$ has minimal percentage error.
\end{itemize}
\end{informaltheorem}

This \href{https://github.com/ATOMSLab/LeanBET/blob/main/BET/KneeSoundness.lean}{theorem} shows that the filtering step has a simple mathematical meaning. The final result is selected from the admissible candidates and satisfies the intended maximal-end-index and minimal-error criteria.

\subsection{Summary of Guarantees}

Taken together, the results of this section show that the LeanBET implementation is connected to a clear mathematical account of the analysis pipeline. Each main stage is supported by a formal statement explaining what the executable step means and what it guarantees when it succeeds.

The linearization result shows that the transformed quantity used in the fit is exactly the one predicted by the nonlinear BET model. The window-enumeration result shows that the search space consists of exactly the valid contiguous fitting windows. The admissibility result (Theorem~A.6) includes monotonicity: it shows that whenever the admissibility checks succeed, the relevant derived sequences are nondecreasing, as required by the formal specification. The regression result shows that a successful fitted line is a least-squares minimizer, and the parameter-extraction result shows that the reported \(n_m\) and \(C\) values agree with the corresponding quantities defined from that certified line. Finally, the knee-selection results show that the final reported result is chosen from the admissible candidates retained according to the intended maximal-end index rule.

Together, these theorems do more than justify isolated pieces of code. They show that the main executable steps of the analysis remain aligned with their mathematical meaning from the beginning to the end. As a result, when the algorithm reports a final BET result, that result is not simply the output of a numerical workflow. It is the output of an executable pipeline whose main stages are tied to machine-checked mathematical guarantees.

\section{Results and Discussion}
\label{sec:discussion}

Across the benchmark set, LeanBET agrees with the BETSI reference implementation to machine precision for 18 of the 19 isotherms. The only exception is UiO-66 \href{https://github.com/ATOMSLab/LeanBET/blob/51a6a8bd0429a893821b57342084fad1fddf7602/Area_results.csv#L18}{result}, for which the two results differ by about \(0.36\,\mathrm{m}^2/\mathrm{g}\), or approximately \(0.03\%\). The full numerical comparison is available in the accompanying repository (see Data Availability).

This level of agreement is important because both implementations rely on IEEE 754 floating-point arithmetic. In that setting, agreement to machine precision across nearly the entire test set provides strong evidence that the Lean implementation faithfully reproduces the intended computational workflow while also supporting formal mathematical guarantees.

The reason for the small difference observed for UiO-66 eludes us, as of this writing. Despite careful comparison, we did not determine why that case differs while the others agree to machine precision. The deviation is much greater than machine precision, but much smaller than typical experimental uncertainties. At present, we simply report the discrepancy as an isolated case and note the otherwise close agreement between BETSI and LeanBET.

Implementing this workflow in Lean was not trivial, but the availability of IEEE 754 floating-point execution together with comparison against a benchmark test suite made it possible to check the implementation while developing the formal results around it. 

As we developed this code, we developed floating-point functions to run tests against BETSI in parallel with Real-valued functions, for exploring derivations and proofs. These were only merged into polymorphic functions at a much later stage. We do not recommend starting a new project with polymorphic functions; writing code for proofs and programs is simply more ergonomic without the typeclass complications. 

These efforts were started by two undergraduate students, one who had specialized in studying theorem proving in Lean, and one who had studied functional programming in Lean. A graduate student who studied both ultimately refactored the separate implementations to create a polymorphic artifact, thus linking the proofs to the program, while expanding the specifications and fixing implementation details that are not represented by the criteria. For instance, BETSI uses PCHIP (Piecewise Cubic Hermite Interpolating Polynomial) interpolation, and our original implementation used linear interpolation, a detail which didn't violate specifications, though it did break agreement with BETSI when different windows were selected. We thus see tests as a critical part of the development process, augmenting the specifications. To put it another way: even with a suite of theorems specifying the derivation of the theory and numerous constraints the implementation must satisfy, there exist multiple variations of the software that would satisfy the proofs; in our case, these were only eliminated by comparison with benchmark tests.

\section{Conclusion}
\label{sec:conclusion}

We have presented LeanBET, a formally verified implementation of BET-style surface area analysis in Lean~4. The development combines executable floating-point computation with proof-oriented specifications over the real numbers, allowing the main steps of the workflow to be connected to explicit mathematical statements. These include the BET linearization, window enumeration, monotonicity checks, least-squares regression, BET parameter extraction, and final knee-based selection.

When compared against the BETSI reference implementation on a benchmark set of adsorption isotherms, LeanBET agrees to machine precision in 18 of the 19 cases. The only exception is UiO-66, where a small discrepancy remains, but this does not affect the overall conclusion of very strong agreement across the benchmark set.

More broadly, this work shows that a practically useful scientific-computing workflow can be translated into Lean without sacrificing numerical agreement. In this setting, formal verification does not replace numerical computation; it helps explain and justify it. By linking the executable pipeline to machine-checked mathematical statements, LeanBET provides a clearer basis for correctness, reproducibility, and trust in BET surface area analysis.

\section*{Acknowledgements} 
This material is based on work supported by the National Science Foundation (NSF) CAREER Award \#2236769.

\section*{Declaration of Interest}
The authors declare no competing financial or personal interests that could influence the work reported in this paper.

\section*{Data Availability}
All code, proofs, and benchmark files are on the \href{https://github.com/ATOMSLab/LeanBET}{ATOMS Lab Github}. 

\section*{ORCID}

\noindent
Ejike D. Ugwuanyi \href{https://orcid.org/0009-0006-4335-5428}{\texttt{https://orcid.org/0009-0006-4335-5428}} \\
Colin T. Jones \href{https://orcid.org/0009-0007-9013-8036}{\texttt{https://orcid.org/0009-0007-9013-8036}} \\
John Velkey \href{https://orcid.org/0000-0001-5156-7451}{\texttt{https://orcid.org/0000-0001-5156-7451}} \\
Tyler R. Josephson \href{https://orcid.org/0000-0002-0100-0227}{\texttt{https://orcid.org/0000-0002-0100-0227}}

\bibliographystyle{unsrturl}
\begin{spacing}{0.5}
\bibliography{references.bib}
\end{spacing}

\clearpage

\end{document}